\documentclass[runningheads]{llncs}

\usepackage[T1]{fontenc}
\usepackage{graphicx}
\usepackage{amsmath,amssymb,amsfonts}
\usepackage{booktabs}
\usepackage{multirow}
\usepackage{array}
\usepackage{makecell}
\usepackage{xcolor}
\usepackage{tikz}
\usepackage{adjustbox}
\usepackage{url}
\usepackage{microtype}
\usetikzlibrary{arrows.meta,positioning,fit,calc,shapes.geometric,backgrounds}
\usepackage{graphicx}

\newcommand{\method}{STEG-OVR}
\newcommand{\base}{Frame-OVR}

\newcommand{\AP}{\mathrm{AP}}
\newcommand{\IoU}{\mathrm{IoU}}
\newcommand{\tIoU}{\mathrm{tIoU}}
\newcommand{\cosim}{\mathrm{cos}}
\newcommand{\expTableFont}{\scriptsize}
\newcommand{\expTableSetup}{\expTableFont\setlength{\tabcolsep}{3pt}\renewcommand{\arraystretch}{1.05}}

\begin{document}

\title{Structured Spatio-Temporal Evidence Graphs for Open-Vocabulary Object Retrieval in Videos}
\titlerunning{Structured Evidence Graphs for Open-Vocabulary Retrieval}


\author{
Jingdan Wang\inst{1} \and
Chuanwen Li\inst{1}\thanks{Corresponding author.}
}

\authorrunning{J. Wang and C. Li}

\institute{
School of Computer Science and Engineering, Northeastern University,
Shenyang, China\\
\email{wangjingdan@stumail.neu.edu.cn}\\
\email{lichuanwen@mail.neu.edu.cn}
}

\maketitle

\begin{abstract}
Open-vocabulary object retrieval in videos requires answering free-form object queries under bounded query-time cost. Existing index-based systems typically store independent frame-level regions and retrieve them with vision--language similarity, which is effective for appearance queries but mismatched with predicates whose evidence is temporal or relational, such as stopped state, scene-region occupancy, persistence, and object interactions. We identify this gap as an evidence-unit mismatch: the query is expressed over tracklets or object tuples, while the index stores isolated boxes. To address it, we propose \method, a structured spatio-temporal evidence graph for open-vocabulary object retrieval. \method{} represents persistent objects as tracklet nodes and temporally compatible object pairs as relation edges, storing appearance, motion, scene occupancy, relative geometry, velocity compatibility, and symbolic relation evidence. A query is decomposed into entity, state, scene, temporal, and relation slots, which activate only the corresponding retrieval channels before soft score fusion and fixed-budget consistency verification. Diagnostic experiments on three object-centric settings show AP improvements from 0.0882 to 0.2073 on Beach, from 0.0834 to 0.1505 on Shibuya, and from 0.701 to 0.743 in a LOVO-style comparison. The gains are strongest for stopped-state and scene/region-occupancy queries, while sustained relations remain sensitive to tracking continuity and predicate calibration.
\end{abstract}

\section{Introduction}

Open-vocabulary object retrieval aims to locate object instances in video collections from free-form textual queries. In contrast to closed-vocabulary video analytics, where categories and predicates are specified before deployment, open-vocabulary retrieval must support long-tail object names, attributes, and contextual descriptions that may be unknown when the archive is indexed. A practical design is to decouple expensive visual computation from online querying: videos are processed offline, region embeddings are extracted by vision--language models, and online queries are answered through approximate nearest-neighbor search and reranking~\cite{radford2021clip,minderer2022owlvit,johnson2019faiss}. LOVO follows this index-based philosophy and shows that compact visual embeddings can support efficient complex object queries in large video collections~\cite{liu2025lovo}.

This design, however, implicitly assumes that the evidence needed by a query can be represented by independent frame-level regions. The assumption is reasonable for appearance-centric queries such as ``a red car'' or ``a person with a backpack''. It becomes problematic for queries such as ``a stopped object'', ``an object in the crosswalk'', ``two objects side by side'', or ``a vehicle following another vehicle''. These predicates are not properties of a single crop alone. They depend on velocity over time, occupancy of a scene region, temporal persistence, or the geometry and motion compatibility between two objects.

We call this limitation an \emph{evidence-unit mismatch}. The semantic unit of the query is a tracklet or a tuple of tracklets, whereas the indexed unit is usually a static detection box. One possible remedy is to run tracking, relation reasoning, or video-language inference at query time, but this weakens the central benefit of index-based retrieval. Our key observation is that the offline index should store evidence at the same granularity as the predicates that users ask about. Appearance can remain open-vocabulary and vector-based, but state, scene, temporal, and relation evidence should be materialized over tracklets and tracklet pairs.

Based on this observation, we propose \method, a structured spatio-temporal evidence graph for open-vocabulary object retrieval. The graph contains tracklet nodes with appearance, motion, duration, trajectory, and scene-occupancy descriptors, and relation edges with temporal overlap, relative geometry, velocity alignment, and symbolic relation labels. At query time, a natural-language query is decomposed into entity, attribute, motion, scene, temporal, and relation slots. These slots activate only the required retrieval channels, allowing candidate generation through vector search, symbolic access buckets, relation-edge joins, soft score fusion, and fixed-budget consistency verification.

This formulation preserves the offline/online separation of LOVO-style retrieval while changing the semantics of the indexed evidence. The method does not replace the underlying open-vocabulary detector or visual encoder; it reorganizes their outputs into predicate-aligned retrieval units. The contributions are threefold. First, we formulate open-vocabulary video object retrieval as structured evidence ranking over tracklets and tracklet tuples, and identify evidence-unit mismatch as a key limitation of frame-level region indexes. Second, we propose \method, a hybrid graph index that materializes tracklet nodes, relation edges, dual-rate sampling evidence, query-channel decomposition, multi-channel scoring, and bounded late consistency checking. Third, we provide diagnostic experiments across three object-centric settings, showing consistent improvements especially for stopped-state and scene/region-occupancy queries, and analyze failure cases for sustained relations and coarse relative regions.

\section{Related Work}

\paragraph{Index-based video analytics.}
Systems such as NoScope, Focus, BlazeIt, OTIF, Seiden, and ExSample reduce online video-query cost by preprocessing videos into indexes, materialized views, specialized models, or adaptive samples~\cite{kang2017noscope,hsieh2018focus,kang2019blazeit,bastani2022otif,bang2023seiden,moll2022exsample}. They show the importance of offline computation for practical video search, but mostly target predefined predicates or fixed categories. \method{} follows the same efficiency principle while targeting open-vocabulary object retrieval and storing tracklet-relation evidence.

\paragraph{LOVO-style open-vocabulary object retrieval.}
LOVO addresses efficient complex object query in videos through one-time feature extraction, compact key-frame embeddings, inverted multi-index storage, and cross-modal reranking~\cite{liu2025lovo}. Its strength is the separation between offline visual computation and online query processing. However, the searchable evidence is still primarily frame-level object evidence, so predicates such as stopped state, scene occupancy, and persistent relations must be approximated from isolated regions or handled by reranking. \method{} keeps the same efficiency principle but changes the indexed evidence unit from boxes to tracklets and temporally compatible tracklet pairs.

\paragraph{Open-vocabulary visual recognition.}
Vision--language pretraining enables direct comparison between text and visual regions~\cite{radford2021clip}. Open-vocabulary detectors, including OWL-ViT, GLIP, and Grounding DINO, extend localization to unseen object names and referring expressions~\cite{minderer2022owlvit,li2022glip,liu2024groundingdino}. These models provide appearance evidence, but independent frame application does not encode stopped state, trajectory interaction, or relation persistence. \method{} treats open-vocabulary recognition as an evidence extractor and adds temporal and relational structure on top.

\paragraph{Tracking and language-based video retrieval.}
Tracking-by-detection methods link object detections into trajectories; ByteTrack shows that associating low-score detections can improve track continuity~\cite{zhang2022bytetrack}. Natural-language moment retrieval and video grounding show that textual queries often describe temporally extended events~\cite{lei2021qvhighlights,liu2022umt,yu2019activitynetqa}. \method{} focuses on an object-centric regime: the output is a tracklet or a tuple of related tracklets, and online computation is bounded by an offline evidence graph.

\section{Problem Definition}

Let $\mathcal{V}=\{v_i\}_{i=1}^{M}$ be a video collection. Each video $v_i$ is a sequence of frames $v_i=\{f_t^i\}_{t=1}^{T_i}$. An open-vocabulary localizer produces a set of detections at sampled frames:
\begin{equation}
    o_t^j=(b_t^j,e_t^j,c_t^j,s_t^j),
\end{equation}
where $b_t^j=(x_t^j,y_t^j,w_t^j,h_t^j)$ is a bounding box, $e_t^j\in\mathbb{R}^{d}$ is the region embedding, $c_t^j$ is an optional open-vocabulary pseudo-category, and $s_t^j\in[0,1]$ is the localization confidence. The spatial overlap between two boxes is measured by intersection over union,
\begin{equation}
    \IoU(b,b')=\frac{|b\cap b'|}{|b\cup b'|},
\end{equation}
where $|\cdot|$ denotes image area. We use this definition both for detection association and for evaluating spatial consistency with scene regions.

A tracklet $\tau_m$ represents one physical object over a temporal interval:
\begin{equation}
    \tau_m=\{o_t^{a_t}\mid t_s^m\leq t\leq t_e^m\},\qquad
    I_m=[t_s^m,t_e^m].
\end{equation}
The index is an evidence graph
\begin{equation}
    \mathcal{G}_{\mathcal{V}}=(\mathcal{T},\mathcal{E},\mathcal{X},\mathcal{R}),
\end{equation}
where $\mathcal{T}$ is the set of tracklets, $\mathcal{E}\subseteq\mathcal{T}\times\mathcal{T}$ is the set of relation edges between temporally compatible tracklets, $\mathcal{X}$ stores node descriptors, and $\mathcal{R}$ stores edge descriptors. A natural-language query $q$ is decomposed into a query graph
\begin{equation}
    G_q=(\mathcal{E}_q,\mathcal{A}_q,\mathcal{M}_q,\mathcal{L}_q,\mathcal{S}_q,\mathcal{T}_q),
\end{equation}
where $\mathcal{E}_q$ denotes entities, $\mathcal{A}_q$ attributes, $\mathcal{M}_q$ motion or state predicates, $\mathcal{L}_q$ pairwise relations, $\mathcal{S}_q$ scene or relative-region constraints, and $\mathcal{T}_q$ temporal constraints. The output is a ranked list of candidates
\begin{equation}
    c\in\mathcal{C}_q=\bigcup_{r=1}^{r_{\max}}\{(\tau_1,\ldots,\tau_r)\in\mathcal{T}^{r}\mid \tau_i\neq\tau_j\;\mathrm{for}\;i\neq j\},
\end{equation}
where $r$ is the number of object entities required by the query. A single-object query returns one tracklet; a relation query returns a tuple of tracklets.

\section{Method}

\subsection{Overview}

\begin{figure}[t]
\centering
\includegraphics[width=\textwidth, trim=10 50 10 40, clip]{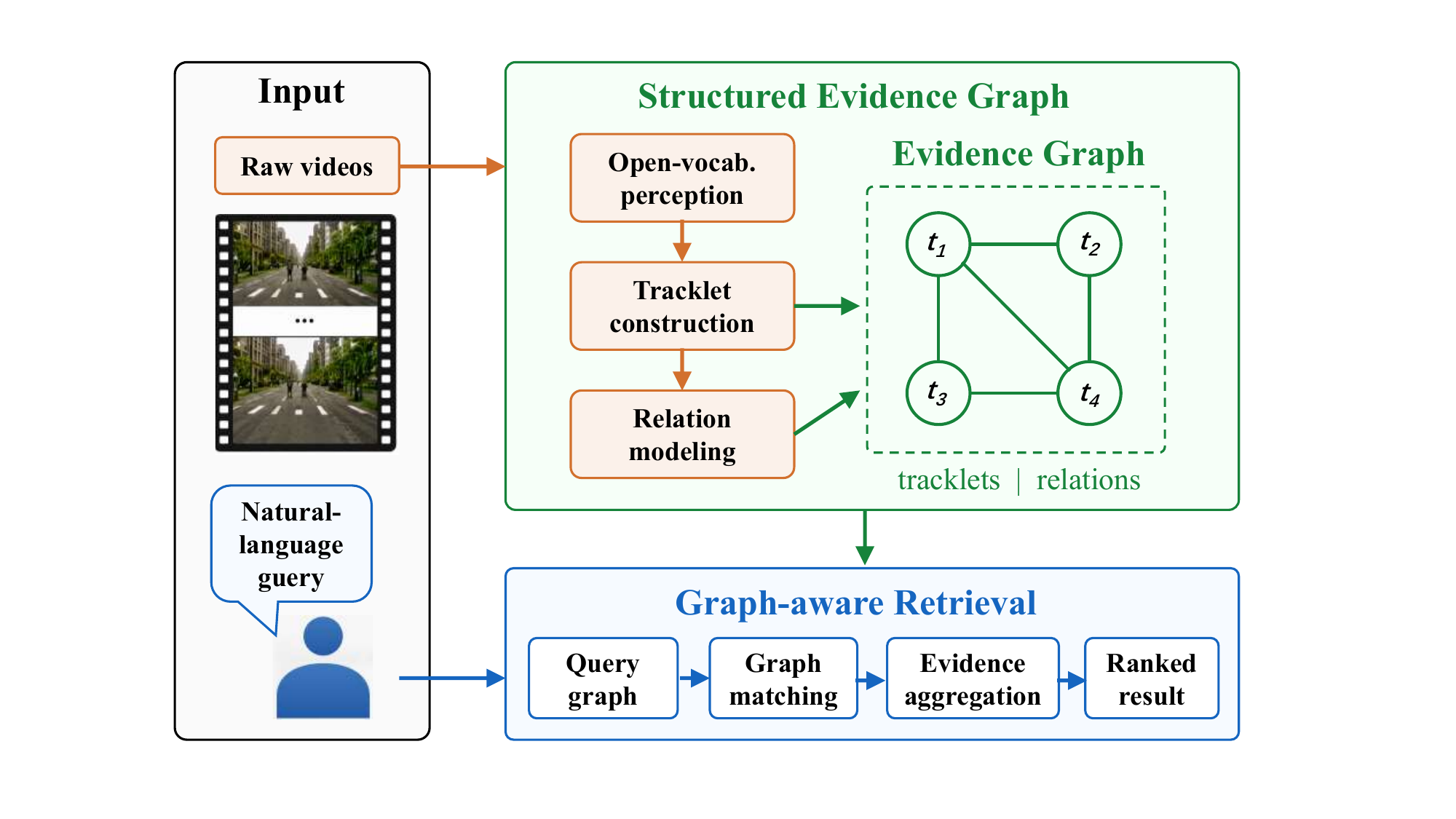}
\caption{Core idea of \method. Videos are converted into a structured evidence graph through open-vocabulary perception, tracklet construction, and relation modeling. Graph-aware retrieval maps a natural-language query into a query graph, matches it against the structured evidence, aggregates evidence, and returns ranked results.}
\label{fig:framework}
\end{figure}

Figure~\ref{fig:framework} gives the overall framework. \method{} is designed to align each query predicate with an indexed evidence unit. Appearance predicates are matched against open-vocabulary region features, state and scene predicates are evaluated on tracklet-level descriptors, and relation predicates are matched against pre-materialized tracklet-pair edges. The offline stage builds this hybrid graph index from dual-rate video samples. The online stage decomposes a query into a structured graph and activates only the evidence channels needed by that query. The key design principle is that temporal and relational structure is materialized offline, while online reasoning is limited to indexed lookup, graph joins, score fusion, and top-$K$ verification.

\subsection{Design Requirements}

\method{} is designed around four requirements. \emph{Evidence granularity} requires the indexed unit to match the query predicate: appearance can be represented by a region embedding, but stopped state, duration, scene occupancy, and relations require temporal aggregation. \emph{Offline materialization} requires reusable visual, motion, scene, and relation evidence to be computed before a query arrives, so query time is not dominated by tracking or pair construction. \emph{Channel selectivity} requires a query to activate only the channels it needs, preventing relation or scene reasoning from being applied to pure appearance queries. \emph{Compatibility} requires reuse of standard open-vocabulary detectors, tracking outputs, and approximate nearest-neighbor indexes. These requirements lead to a hybrid design: node evidence is used for single-object queries, edge evidence for relation queries, and symbolic buckets are used only as high-recall access paths. Final ranking remains soft and multi-channel.

Operationally, \method{} samples key frames for open-vocabulary appearance evidence and denser tracking frames for temporal continuity, links detections into tracklets, computes node descriptors, and materializes relation edges only for temporally compatible tracklet pairs. At query time, entity and attribute slots query the appearance index, state and scene slots access high-recall buckets, and relation slots join entity candidates through precomputed edges before channel-normalized fusion and fixed-budget verification.

\subsection{Dual-Rate Evidence Acquisition}

A single sampling rate creates an undesirable tradeoff. Sparse sampling is efficient for appearance indexing but damages tracking and motion estimation; dense sampling improves temporal continuity but makes open-vocabulary localization and embedding extraction unnecessarily expensive. We therefore use two synchronized sampling streams. Let $\rho_k$ be the key-frame rate and $\rho_t$ the tracking-frame rate, with $\rho_k<\rho_t$. In the implementation reflected by the experiments, $\rho_k=1$ fps and $\rho_t=5$ fps. For a video $v$ with native frame rate $r_v$ and duration $T_v$, the sampled set at rate $\rho$ is
\begin{equation}
    \mathcal{F}_{\rho}(v)=\{f_{\lfloor n r_v/\rho\rfloor}\mid n=0,\ldots,\lfloor \rho T_v\rfloor-1\}.
\end{equation}
The key-frame and tracking-frame sets are then
\begin{equation}
    \mathcal{F}_k(v)=\mathcal{F}_{\rho_k}(v),\qquad
    \mathcal{F}_t(v)=\mathcal{F}_{\rho_t}(v).
\end{equation}
Key frames are used to compute expensive open-vocabulary embeddings $e_t^j$ and build the appearance index. Tracking frames are used to maintain short-range association, estimate velocity, and recover temporal relations. This separates semantic indexing cost from temporal sampling density.

For consecutive tracking frames, detections are linked by minimizing an association cost
\begin{equation}
\begin{aligned}
    D(o_t^i,o_{t+\Delta}^{j})={}&\lambda_{\IoU}\bigl(1-\IoU(b_t^i,b_{t+\Delta}^{j})\bigr) \\
    &+\lambda_c\frac{\|\mu(b_t^i)-\mu(b_{t+\Delta}^{j})\|_2}{\Delta}
    +\lambda_e\chi_{ij}^{t,t+\Delta}\bigl(1-\cosim(e_t^i,e_{t+\Delta}^{j})\bigr).
\end{aligned}
\end{equation}
where $\mu(b)$ is the box center and $\chi_{ij}^{t,t+\Delta}\in\{0,1\}$ indicates whether both detections have valid appearance embeddings. Matches are solved by bipartite assignment with a maximum-cost rejection threshold; unmatched detections start new tracklets, and unmatched active tracklets are retained for a short grace period. This cost explains how the dual-rate design remains compatible with a key-frame embedding index such as LOVO: dense frames refine tracks even if only key frames carry full open-vocabulary features.

\subsection{Predicate-Aligned Tracklet Evidence}

For each tracklet $\tau$, \method{} builds a node descriptor $x_{\tau}$ composed of appearance, motion, trajectory, state, and scene evidence:
\begin{equation}
    x_{\tau}=\bigl[a_{\tau},m_{\tau},g_{\tau},p_{\mathrm{state}}(\tau),p_{\mathrm{scene}}(\tau)\bigr].
\end{equation}
The appearance descriptor is a confidence-weighted and normalized average of key-frame embeddings,
\begin{equation}
    a_{\tau}=\frac{\sum_{t\in I_\tau\cap\mathcal{F}_k} s_t e_t}{\left\|\sum_{t\in I_\tau\cap\mathcal{F}_k} s_t e_t\right\|_2+\epsilon}.
\end{equation}
This preserves the open-vocabulary matching ability of CLIP/OWL-ViT-style features while reducing tracklet evidence to one searchable vector.

Motion evidence is computed from the center trajectory $c_t=\mu(b_t)$ after normalizing image coordinates to $[0,1]^2$:
\begin{equation}
    u_{\ell}=\frac{c_{t_{\ell+1}}-c_{t_{\ell}}}{t_{\ell+1}-t_{\ell}},\qquad
    \bar{u}_{\tau}=\frac{1}{L_\tau-1}\sum_{\ell=1}^{L_\tau-1}\|u_{\ell}\|_2 .
\end{equation}
For $L_\tau=1$, $\bar{u}_{\tau}$ is set to $+\infty$ so that a single detection cannot be classified as stopped. A soft stopped-state score is defined as
\begin{equation}
    p_{\mathrm{stop}}(\tau)=\sigma\left(\alpha(\delta_u-\bar{u}_{\tau})\right),
\end{equation}
where $\delta_u$ is a speed threshold and $\sigma$ is the sigmoid function. This formulation avoids a brittle binary decision: slow tracklets receive high stopped scores, while noisy but nearly stationary tracks are not immediately discarded.

Scene and relative-region evidence are stored as soft occupancy rather than a single hard label. For a predefined region $r$ with image mask or polygon $\Omega_r$, the occupancy is
\begin{equation}
    p_r(\tau)=\frac{1}{|I_\tau|}\sum_{t\in I_\tau}\frac{|b_t\cap\Omega_r|}{|b_t|}.
\end{equation}
This descriptor supports absolute scene queries such as ``in the crosswalk'' and relative-region queries such as ``in the left region''. Candidate generation uses a conservative bucket $B_s(r;\eta_r)=\{\tau\mid p_r(\tau)\geq\eta_r\}$, while the continuous score $p_r(\tau)$ is retained for score fusion. This distinction is important: if a query region is coarse or a tracklet crosses a boundary, an overly high $\eta_r$ can over-filter relevant objects.

\subsection{Temporally Compatible Relation Evidence}

For two tracklets $\tau_i$ and $\tau_j$, an edge is created only when their temporal intervals overlap sufficiently. Temporal overlap is measured by
\begin{equation}
    \tIoU(\tau_i,\tau_j)=\frac{|I_i\cap I_j|}{|I_i\cup I_j|}.
\end{equation}
When $\tIoU(\tau_i,\tau_j)>\eta_t$, the edge descriptor stores relative geometry and motion statistics over the shared interval $I_{ij}=I_i\cap I_j$:
\begin{equation}
    r_{ij}=\left[\overline{\Delta c}_{ij},\overline{d}_{ij},\overline{\kappa}_{ij},\tIoU(\tau_i,\tau_j),\ell_{ij}\right].
\end{equation}
Here $\overline{\Delta c}_{ij}$ is the normalized relative center displacement, $\overline{d}_{ij}$ is the mean object distance, $\overline{\kappa}_{ij}$ is the mean velocity alignment, and $\ell_{ij}$ is a symbolic relation label such as \emph{near}, \emph{left-of}, \emph{right-of}, \emph{side-by-side}, or \emph{following}. A following score, for example, can be written as
\begin{equation}
    \psi_{\mathrm{follow}}(i,j)=\sigma\left(\gamma_1\overline{\kappa}_{ij}-\gamma_2\overline{d}_{ij}+\gamma_3\tIoU(\tau_i,\tau_j)\right).
\end{equation}
This edge representation is deliberately hybrid. Vector terms provide graded similarity, while symbolic labels enable efficient filtering. Thus relation reasoning does not require scanning all object pairs online; candidate pairs are pre-materialized as graph edges.

\subsection{Hybrid Graph Index Layout}

After node and edge construction, the evidence graph is stored in a hybrid index
\begin{equation}
    \mathcal{I}=(I_a,I_m,I_s,I_t,I_r),
\end{equation}
where $I_a$ is an approximate nearest-neighbor index over appearance descriptors, $I_m$ stores motion and state buckets, $I_s$ stores scene-region buckets, $I_t$ stores duration or temporal-interval buckets, and $I_r$ stores relation-label lists with pointers to edge descriptors. The index also stores back-pointers from each tracklet to its original boxes and timestamps. These pointers are not scanned during candidate generation; they are accessed only for visualization and bounded verification.

This layout separates dense vector search from sparse symbolic access. Appearance matching provides open-vocabulary recall, while motion, scene, temporal, and relation buckets restrict the candidate set when the query explicitly requests such evidence. The design is therefore sublinear in raw detections at query time and avoids enumerating all object pairs for relation queries.

\subsection{Query-Time Cost and Evidence Semantics}

The graph index is intended to change the semantics of retrieved evidence without changing the online cost model of index-based retrieval. Let $N$ be the number of frame-level detections, $T$ the number of tracklets, and $E$ the number of pre-materialized relation edges. A frame-level baseline retrieves from $N$ visual regions. In contrast, \method{} retrieves from $T$ tracklet nodes for single-object queries and from the relevant subset of $E$ edges for relation queries. Since detections belonging to the same physical object are collapsed into one tracklet, typically $T\ll N$ for videos with persistent objects. Relation candidates are not produced by online pair enumeration over $T^2$ pairs; only temporally compatible edges are stored and queried.

For a single-object query, the dominant online operations are an approximate nearest-neighbor lookup in $I_a$, a small number of bucket accesses in $I_m$, $I_s$, or $I_t$, and score fusion over the resulting shortlist. For a two-object relation query, candidate pairs are generated by joining entity shortlists through $I_r$. The cost is therefore governed by the shortlist size and the number of matching indexed edges, not by the raw video length. This property is important for PRICAI-style AI systems settings: the method adds explicit symbolic and geometric evidence, but it keeps expensive visual inference and pair construction offline.

The same structure also clarifies what each retrieved item means. A high score from $I_a$ says that a tracklet looks like the entity phrase; a high motion score says that its trajectory satisfies a state predicate; a high scene score says that it occupies a named or relative region over time; and a high relation score says that two tracklets satisfy a precomputed temporal-geometric relation. This decomposition makes the ranking more auditable than a single cross-modal similarity score and explains the family-level behavior observed in the experiments.

\subsection{Query Graph Decomposition}

A text query $q$ is decomposed into a query graph $G_q$. The parser does not attempt full language understanding; it exposes retrieval slots. Entity and attribute phrases remain open-vocabulary appearance queries, while lexicon/template rules map terms such as \emph{stopped}, \emph{moving}, \emph{near}, \emph{next to}, \emph{following}, scene names, and relative-region words to motion, relation, temporal, and scene slots. Unmatched words stay in the appearance phrase, which avoids discarding useful text when a phrase cannot be mapped to a symbolic slot. We represent the active channel mask as
\begin{equation}
    z_q=(z_a,z_m,z_s,z_r,z_t)\in\{0,1\}^{5},
\end{equation}
corresponding to appearance, motion, scene, relation, and temporal channels. This mask controls which indexes are queried and which score terms are included, so pure appearance queries do not invoke relation verification and relation queries are not reduced to appearance similarity.

\begin{table}[t]
\centering
\caption{Representative query decomposition patterns used in the experiments. Entity and attribute phrases remain open-vocabulary appearance queries; structural predicates use lexicon/template slots.}
\label{tab:parser}
\expTableSetup
\begin{adjustbox}{max width=\textwidth}
\begin{tabular}{llll}
\toprule
Query form & Active slots & Candidate arity & Main evidence channel\\
\midrule
\emph{a stopped object} & entity, state & 1 & tracklet velocity statistics\\
\emph{an object in the crosswalk} & entity, scene & 1 & scene occupancy bucket and score\\
\emph{an object in the bottom region} & entity, relative region & 1 & soft region occupancy\\
\emph{an object next to another object} & two entities, relation & 2 & precomputed relation edge\\
\emph{an object following another object} & two entities, sustained relation & 2 & temporal overlap, distance, velocity alignment\\
\emph{a moving object} & entity, motion & 1 & normalized trajectory speed\\
\bottomrule
\end{tabular}
\end{adjustbox}
\end{table}

\subsection{Candidate Generation}

Given the active channel mask, \method{} first retrieves candidates from each active index and then intersects only the symbolic buckets requested by the query. These bucket intersections are candidate-access operations, not final relevance decisions: thresholds are set conservatively and continuous motion/scene scores remain in the fusion stage. For a single-object query,
\begin{equation}
    \mathcal{C}_{q}^{(1)}=\mathrm{TopK}_a(q)\cap\bigcap_{k\in\{m,s,t\}:z_k(q)=1}B_k(q),
\end{equation}
where $\mathrm{TopK}_a(q)$ is the appearance shortlist and $B_m$, $B_s$, and $B_t$ are motion, scene, and temporal buckets. If a bucket is absent or unreliable for a query, the corresponding predicate is handled as a soft score rather than as a hard access constraint. For a relation query, node candidates are generated for each entity and joined through precomputed graph edges:
\begin{equation}
    \mathcal{C}_{q}^{(2)}=\{(\tau_i,\tau_j)\mid \tau_i\in\mathcal{C}_{q,1}^{(1)},\tau_j\in\mathcal{C}_{q,2}^{(1)},(\tau_i,\tau_j)\in\mathcal{E},\ell_{ij}\sim \ell_q\}.
\end{equation}
Here $\ell_{ij}\sim\ell_q$ denotes compatibility between the indexed relation label and the requested relation. This join avoids enumerating all object pairs online.

\subsection{Channel-Selective Retrieval and Score Fusion}

For a single-tracklet candidate $\tau$, the retrieval score is a sum of active evidence terms rather than a monolithic cross-modal score:
\begin{equation}
    S_{\mathrm{node}}(q,\tau)=\lambda_a(q)\phi_a(q,\tau)+\lambda_m(q)\phi_m(q,\tau)+\lambda_s(q)\phi_s(q,\tau)+\lambda_t(q)\phi_t(q,\tau),
\end{equation}
where $\phi_a$ is text--appearance similarity, $\phi_m$ evaluates motion or state predicates, $\phi_s$ evaluates scene or relative-region membership, and $\phi_t$ evaluates duration constraints. For a relation candidate $(\tau_i,\tau_j)$, we add an edge score
\begin{equation}
    S_{\mathrm{edge}}(q,\tau_i,\tau_j)=\lambda_r(q)\phi_r(q,r_{ij})+\frac{1}{2}S_{\mathrm{node}}(q,\tau_i)+\frac{1}{2}S_{\mathrm{node}}(q,\tau_j).
\end{equation}
The channel weights are query-adaptive:
\begin{equation}
    \lambda_k(q)=\frac{z_k(q)\exp(\theta_k)}{\sum_{\ell}z_\ell(q)\exp(\theta_\ell)+\epsilon},\qquad k\in\{a,m,s,r,t\}.
\end{equation}
The mask $z_k(q)$ removes irrelevant channels, while the learnable or tuned prior $\theta_k$ controls the relative importance of active evidence types. This formulation is more interpretable than a monolithic neural score because each component has a direct retrieval meaning.

After candidate generation, \method{} applies optional fixed-budget temporal consistency scoring to the top-ranked candidates. In the reported experiments, the same late scoring budget is used for the full model and ablations, so the comparison primarily evaluates how the graph index changes the candidate set and ranking before this final check. The scorer loads a short clip around the candidate interval, target object crops, related object crops when present, trajectory tokens, relation tokens, and the query graph. It produces a bounded consistency score
\begin{equation}
    S_{\mathrm{cons}}(q,c)\in[0,1].
\end{equation}
The final score is
\begin{equation}
    S_{\mathrm{final}}(q,c)=\beta S_{\mathrm{retrieval}}(q,c)+(1-\beta)S_{\mathrm{cons}}(q,c).
\end{equation}
This scorer is not used as an additional full-video reasoning model. It is a fixed-budget consistency check over a small candidate set, so its cost scales with the candidate budget rather than with the total video length.

\section{Experiments}

\subsection{Experimental Setup}

We report empirical results across three object-centric evaluation settings. Beach and Shibuya are diagnostic retrieval settings with complete per-query and per-family logs. Each contains 13 queries covering spatial relation, sustained relation, multi-object proximity, temporal action, stopped state, and scene/relative-region constraints. The third setting is an object-centric spatio-temporal benchmark with 10 weakly supervised queries, 16 tracklets, and 643 pairwise relations. It is used as a complementary comparison against the LOVO baseline.

The primary baseline is a frame-level open-vocabulary retrieval system following the same offline indexing principle as LOVO. On Beach and Shibuya, this baseline is denoted \base{} because it ranks isolated frame/object evidence and does not materialize tracklet state, scene occupancy, or relation edges. On the object-centric benchmark, the baseline is denoted LOVO because the available comparison follows the published LOVO-style object retrieval protocol. The comparisons are organized to isolate evidence representation. In each setting, the baseline and \method{} use the same evaluated detections, region embeddings, query text, and metric implementation; \method{} changes how this evidence is aggregated and indexed by adding tracklet nodes, scene/motion descriptors, and relation edges. The bounded consistency scoring budget is fixed across full and ablated variants. For all settings, higher AP, recall, precision, and nDCG are better. AP is macro-averaged over queries. The subsequent per-family, per-query, and time analyses are reported on Beach and Shibuya because these two settings provide complete query-family logs.

\subsection{Metrics and Implementation Details}

For a query $q$, candidates are sorted by $S_{\mathrm{final}}(q,c)$ and compared against annotated relevant tracklets or tracklet tuples. Precision and recall at rank $K$ evaluate top-ranked retrieval quality. Average precision is computed as
\begin{equation}
    \AP(q)=\frac{1}{N_q^+}\sum_{k=1}^{N_q} P_q(k)\mathrm{rel}_q(k),
\end{equation}
where $N_q^+$ is the number of relevant candidates, $P_q(k)$ is precision at rank $k$, and $\mathrm{rel}_q(k)\in\{0,1\}$ indicates whether the candidate at rank $k$ is relevant. We report macro AP by averaging over queries. Ranking quality is also measured by
\begin{equation}
    \mathrm{nDCG}@K(q)=\frac{1}{\mathrm{IDCG}_K(q)}\sum_{k=1}^{K}\frac{2^{\mathrm{rel}_q(k)}-1}{\log_2(k+1)} .
\end{equation}
AP measures full-list ranking, recall measures coverage, precision measures top-ranked correctness, and nDCG rewards placing relevant candidates early.

All comparisons reuse the same visual evidence source within each setting. The frame-level baseline ranks isolated frame/object evidence, while \method{} aggregates the same detections into tracklets and relation edges. Query decomposition is fixed before evaluation and is shared by the full model and the channel ablations. Scene and motion thresholds are used only to construct high-recall buckets; their continuous scores remain in the fusion stage. Unless otherwise stated, the bounded verifier is applied with the same candidate budget across variants, so ablation differences reflect the indexed evidence channels rather than a larger verification budget. Table~\ref{tab:impl} summarizes the implementation choices that affect reproducibility.

\begin{table}[t]
\centering
\caption{Implementation details shared across the reported experiments. Thresholds are used for candidate access; the corresponding continuous scores are still used during fusion.}
\label{tab:impl}
\expTableSetup
\begin{tabular}{p{0.24\textwidth}p{0.68\textwidth}}
\toprule
Component & Setting\\
\midrule
Open-vocabulary evidence & same evaluated detections, region embeddings, and query text for compared methods within each setting\\
Sampling & 1 fps key frames for appearance evidence; 5 fps tracking frames for temporal continuity\\
Tracking & bipartite matching with IoU, center displacement, and gated appearance cost\\
Query parser & open-vocabulary entity/attribute phrases; lexicon/template slots for motion, scene, relation, and time\\
Scene and motion access & conservative symbolic buckets plus continuous occupancy/speed scores in fusion\\
Relation access & pre-materialized temporally overlapping tracklet pairs with symbolic relation labels\\
Fusion weights & fixed query-channel priors normalized over active channels\\
Consistency scoring & same bounded top-ranked candidate budget for full model and ablations\\
Timing & online retrieval plus fixed-budget consistency scoring on Beach/Shibuya query logs; offline indexing excluded\\
\bottomrule
\end{tabular}
\end{table}

\subsection{Main Results on Three Datasets}

\begin{table}[t]
\centering
\caption{Main retrieval comparison across three object-centric evaluation settings. Higher is better.}
\label{tab:main_three}
\expTableSetup
\begin{adjustbox}{max width=\textwidth}
\begin{tabular}{llcccccc}
\toprule
Dataset & Method & AP & R@5 & R@10 & P@10 & nDCG@10 & $\Delta$AP \\
\midrule
\multirow{2}{*}{Beach} & Frame-OVR & 0.0882 & 0.0201 & 0.0447 & 0.3833 & 0.4099 & -- \\
 & \method{} & \textbf{0.2073} & \textbf{0.0819} & \textbf{0.1281} & \textbf{0.5854} & \textbf{0.6319} & +0.1191 \\
\midrule
\multirow{2}{*}{Shibuya} & Frame-OVR & 0.0834 & 0.0179 & 0.0357 & 0.3438 & 0.3378 & -- \\
 & \method{} & \textbf{0.1505} & \textbf{0.0550} & \textbf{0.0850} & \textbf{0.5396} & \textbf{0.5344} & +0.0671 \\
\midrule
\multirow{2}{*}{Object-centric benchmark} & LOVO~\cite{liu2025lovo} & 0.701 & 0.344 & 0.613 & 0.613 & 0.721 & -- \\
 & \method{} & \textbf{0.743} & \textbf{0.353} & \textbf{0.646} & \textbf{0.677} & \textbf{0.750} & +0.042 \\
\bottomrule
\end{tabular}
\end{adjustbox}
\end{table}

Table~\ref{tab:main_three} shows the main retrieval results. On Beach, AP increases from 0.0882 to 0.2073; on Shibuya, AP increases from 0.0834 to 0.1505. These gains are mainly associated with state and scene-constrained queries, where frame-level regions lack the temporal or region-occupancy evidence required by the query. On the object-centric benchmark, \method{} improves AP from 0.701 to 0.743 against LOVO. The gain is smaller, but it is consistent with the view that tracklet and relation evidence can complement frame-level open-vocabulary matching.

The absolute recall values remain modest, especially on Beach and Shibuya, so the results should be read as precision-oriented retrieval improvements rather than complete coverage of all relevant objects. This interpretation is consistent with the top-$K$ analysis below: the strongest effect of the graph index is to move relevant structured candidates earlier in the ranking.

\subsection{Top-$K$ Ranking Behavior on Beach and Shibuya}

\begin{table}[t]
\centering
\caption{Top-$K$ retrieval profile averaged over Beach and Shibuya. The analysis uses the two datasets with complete top-$K$ logs.}
\label{tab:topk}
\expTableSetup
\begin{adjustbox}{max width=\textwidth}
\begin{tabular}{lccccccccc}
\toprule
Method & R@5 & R@10 & R@20 & R@30 & P@5 & P@10 & P@20 & nDCG@10 & nDCG@20\\
\midrule
Frame-OVR & 0.0190 & 0.0402 & 0.0616 & 0.0685 & 0.3812 & 0.3635 & 0.3031 & 0.3739 & 0.3286 \\
\method{} & \textbf{0.0684} & \textbf{0.1066} & \textbf{0.1459} & \textbf{0.1594} & \textbf{0.5625} & \textbf{0.5625} & \textbf{0.4805} & \textbf{0.5832} & \textbf{0.5170} \\
Abs. gain & +0.0494 & +0.0664 & +0.0843 & +0.0909 & +0.1813 & +0.1990 & +0.1774 & +0.2093 & +0.1884 \\
\bottomrule
\end{tabular}
\end{adjustbox}
\end{table}

The early-rank behavior matters because video retrieval systems are usually inspected through the first few returned instances. Table~\ref{tab:topk} shows that \method{} improves both recall and precision in the top ranks. This means that the graph index does not merely retrieve more candidates at large $K$; it also places relevant tracklets earlier in the ranking. The gain in nDCG@10 further indicates that the improvement is not only a filtering effect but also a ranking-quality effect.

\subsection{Query-Family Analysis on Beach and Shibuya}

\begin{table}[t]
\centering
\caption{AP by query family on Beach and Shibuya. Scene constraints include absolute scene regions such as crosswalk/intersection and relative-region queries such as left/right/bottom/center.}
\label{tab:family}
\expTableSetup
\begin{adjustbox}{max width=\textwidth}
\begin{tabular}{lccccccc}
\toprule
\multirow{2}{*}{Query family} & \multicolumn{3}{c}{Beach} & \multicolumn{3}{c}{Shibuya} & \multirow{2}{*}{Avg. $\Delta$}\\
\cmidrule(lr){2-4}\cmidrule(lr){5-7}
& Base & \method{} & $\Delta$ & Base & \method{} & $\Delta$ & \\
\midrule
Spatial relation & 0.0228 & 0.0449 & +0.0221 & 0.0000 & 0.0000 & +0.0000 & +0.0111 \\
Sustained relation & 0.0843 & 0.0890 & +0.0047 & 0.0817 & 0.0659 & -0.0158 & -0.0055 \\
Multi-object proximity & 0.0760 & 0.1082 & +0.0322 & 0.0397 & 0.0504 & +0.0107 & +0.0215 \\
Temporal action & 0.1749 & 0.2111 & +0.0362 & 0.2405 & 0.2161 & -0.0244 & +0.0059 \\
Stopped state & 0.0119 & 0.3103 & +0.2984 & 0.0027 & 0.2231 & +0.2204 & +0.2594 \\
Scene/relative-region constraint & 0.1594 & 0.4805 & +0.3211 & 0.1359 & 0.3476 & +0.2117 & +0.2664 \\
\bottomrule
\end{tabular}
\end{adjustbox}
\end{table}

Table~\ref{tab:family} explains where the overall improvement comes from. The largest gains appear in stopped-state and scene/relative-region queries. This is consistent with the method design: stopped-state queries depend on velocity statistics aggregated over tracklets, while scene queries depend on region occupancy accumulated over time. Multi-object proximity also improves because pairwise edges reduce the ambiguity of independently retrieved objects.

Sustained relation and temporal action are not the main source of improvement. These families depend strongly on tracking quality and predicate calibration. If a trajectory is fragmented, or if ``following'' is expressed by a loose distance rule rather than a learned relation score, structured evidence can introduce false negatives. Thus the current graph is most reliable for tracklet state, scene occupancy, and region constraints, while sustained relation reasoning requires better edge calibration.

\subsection{Component Ablation}

\begin{table}[t]
\centering
\caption{Component ablation on Beach and Shibuya. R10, P10, and N10 denote Recall@10, Precision@10, and nDCG@10; Time is wall-clock evaluation time in seconds.}
\label{tab:ablation_two}
\expTableSetup
\begin{adjustbox}{max width=\textwidth}
\begin{tabular}{lccccc|ccccc}
\toprule
\multirow{2}{*}{Variant} & \multicolumn{5}{c|}{Beach} & \multicolumn{5}{c}{Shibuya}\\
\cmidrule(lr){2-6}\cmidrule(lr){7-11}
& AP & R10 & P10 & N10 & Time & AP & R10 & P10 & N10 & Time\\
\midrule
Full \method{} & \textbf{0.2073} & \textbf{0.1281} & \textbf{0.5854} & \textbf{0.6319} & 193.0 & \textbf{0.1505} & \textbf{0.0850} & \textbf{0.5396} & \textbf{0.5344} & 151.9 \\
w/o relation channel & 0.1883 & 0.1154 & 0.5600 & 0.6104 & 191.6 & 0.1446 & 0.0804 & 0.5300 & 0.5207 & 151.3 \\
w/o motion channel & 0.1766 & 0.1095 & 0.5514 & 0.6209 & 192.1 & 0.1350 & 0.0742 & 0.5167 & 0.5038 & 152.0 \\
w/o scene channel & 0.1427 & 0.0857 & 0.5229 & 0.5694 & 193.4 & 0.1052 & 0.0573 & 0.4937 & 0.4896 & 152.7 \\
Frame-OVR & 0.0882 & 0.0447 & 0.3833 & 0.4099 & 181.5 & 0.0834 & 0.0357 & 0.3438 & 0.3378 & 153.9 \\
\bottomrule
\end{tabular}
\end{adjustbox}
\end{table}

Table~\ref{tab:ablation_two} indicates that scene evidence is the most important component in the two fully logged datasets, followed by motion and relation evidence. This ordering is consistent across Beach and Shibuya and matches the query families: scene evidence supports road, sidewalk, crosswalk, intersection, and relative-region queries; motion supports stopped and moving predicates; relation edges provide a smaller but positive signal for object-pair queries. The pattern supports the interpretation that \method{} is not merely a larger reranker, but a combination of distinct indexed evidence channels.

The object-centric benchmark provides a complementary ablation: full \method{} obtains AP/R@10/nDCG@10 of $0.743/0.646/0.750$, while removing query decomposition gives $0.701/0.613/0.721$, removing the motion channel gives $0.688/0.635/0.687$, and removing the scene channel gives $0.709/0.622/0.719$. This again supports the claim that the gains arise from semantically specialized evidence rather than from merely adding parameters or reranking every candidate.

\subsection{Per-Query Diagnostics and Region Calibration}

\begin{table}[t]
\centering
\caption{Representative per-query AP diagnostics. Positive $\Delta$ indicates improvement over the frame-level baseline; lower values highlight calibration-sensitive structured predicates.}
\label{tab:perquery}
\expTableSetup
\begin{tabular}{p{0.25\textwidth}p{0.15\textwidth}p{0.09\textwidth}p{0.10\textwidth}p{0.27\textwidth}}
\toprule
Query & Family & Beach $\Delta$ & Shibuya $\Delta$ & Interpretation\\
\midrule
an object in the crosswalk & Scene constraint & +0.9736 & +0.9063 & strong gain from explicit region occupancy\\
an object in the intersection & Scene constraint & +1.0000 & +0.8085 & strong gain from scene-region filtering\\
a stopped object & Stopped state & +0.2985 & +0.2204 & velocity aggregation resolves static-state query\\
an object next to another object & Multi-object & +0.0322 & +0.0106 & relation edges improve object-pair retrieval\\
an object following another object & Sustained relation & +0.0047 & -0.0158 & sensitive to trajectory fragmentation\\
a moving object & Temporal action & +0.0362 & -0.0243 & affected by motion threshold calibration\\
an object in the bottom region & Relative region & -0.1441 & -0.0360 & coarse region bin can over-filter\\
an object in the center of road & Relative/scene region & -0.1106 & -0.1871 & boundary crossing and ambiguous annotations\\
\bottomrule
\end{tabular}
\end{table}

The lower values in Table~\ref{tab:perquery} identify predicates that need careful calibration. They mostly occur in coarse relative-region queries such as bottom, left/right, and center-of-road. These predicates are harder than named scene regions because their boundaries are not semantically stable: an object can cross from center to bottom during a tracklet, camera perspective changes the apparent region, and annotation protocols may label the object by a representative frame rather than by full temporal occupancy. In the current implementation, a hard threshold on $p_r(\tau)$ may therefore penalize a relevant tracklet. The strong gains on crosswalk, intersection, sidewalk, road, and stopped-state queries show that structured scene and temporal evidence is beneficial when the region semantics are well defined, while coarse relative regions call for softer calibration.

\subsection{Efficiency}

\begin{table}[t]
\centering
\caption{Online time cost on Beach and Shibuya. Timing covers retrieval plus fixed-budget consistency scoring on the query logs; offline detection, tracking, and index construction are excluded. AP/s and N10/s are multiplied by $10^3$ for readability.}
\label{tab:time}
\expTableSetup
\begin{adjustbox}{max width=\textwidth}
\begin{tabular}{llcccccc}
\toprule
Dataset & Method & Q & Time (s) & s/query & $10^3$ AP/s & $10^3$ N10/s & nDCG@10\\
\midrule
Beach & Frame-OVR & 13 & 181.5 & 13.96 & 0.486 & 2.258 & 0.4099 \\
Beach & \method{} & 13 & 193.0 & 14.85 & 1.074 & 3.274 & 0.6319 \\
Shibuya & Frame-OVR & 13 & 153.9 & 11.84 & 0.542 & 2.195 & 0.3378 \\
Shibuya & \method{} & 13 & 151.9 & 11.68 & 0.991 & 3.518 & 0.5344 \\
Average & Frame-OVR & 13 & 167.7 & 12.90 & 0.512 & 2.229 & 0.3739 \\
Average & \method{} & 13 & 172.4 & 13.27 & 1.037 & 3.382 & 0.5832 \\
\bottomrule
\end{tabular}
\end{adjustbox}
\end{table}

Table~\ref{tab:time} reports online wall-clock evaluation time on the Beach and Shibuya query logs. On average, time increases from 167.7s to 172.4s. Because offline detection, tracking, graph construction, and index building are excluded, the table should be read as online retrieval plus fixed-budget consistency scoring. After graph construction, adding tracklet descriptors, relation edges, and scene memberships does not substantially increase measured online time in these logs.

\section{Discussion}

The results support a bounded but useful conclusion: open-vocabulary retrieval benefits when indexed evidence matches the query predicate. State and scene predicates align well with tracklet velocity and occupancy descriptors, while relation predicates require reliable temporal overlap, trajectory continuity, and calibrated distance or velocity thresholds. Coarse relative-region phrases are also sensitive to camera perspective and boundary crossing. These observations point to learned relation/region calibration and adaptive access thresholds that keep symbolic buckets high-recall while leaving final relevance to soft fusion and bounded verification.

\section{Conclusion}

This paper presented \method, a structured spatio-temporal evidence graph for open-vocabulary object retrieval in videos. The method builds a hybrid graph index over tracklets and relation edges, parses queries into structured evidence requirements, and combines vector retrieval with symbolic filtering and fixed-budget consistency scoring. Diagnostic experiments across three object-centric settings show AP improvements, especially for stopped-state and scene/region-occupancy queries. The main conclusion is that open-vocabulary appearance retrieval benefits from structured indexed evidence when queries contain explicit state, scene, or selected relation predicates.

\end{document}